\documentclass[letterpaper,twocolumn,10pt]{article}
\usepackage{usenix,epsfig,endnotes}

\usepackage[T1]{fontenc}% optional T1 font encoding

\usepackage[cmintegrals]{newtxmath}
\usepackage{url}
\newcommand\Mark[1]{\textsuperscript{#1}}

\usepackage{hyperref}
\usepackage{tcolorbox}

\usepackage{tabularx,adjustbox}
\usepackage{booktabs}
\usepackage{graphicx}  
\usepackage{algorithm,algpseudocode} 
\usepackage{longfbox}
\usepackage{subcaption}

\usepackage{amsmath, amscd, mathrsfs}

\usepackage{multirow}

\usepackage{array}
\usepackage{xurl}

\usepackage{pifont}% 

\newcommand{\raf}[1]{(\ref{#1})}

\makeatletter
\newcommand{\vast}{\bBigg@{4}}
\newcommand{\Vast}{\bBigg@{5}}
\newcommand{\immense}{\bBigg@{6}}
\newcommand{\Immense}{\bBigg@{7}}
\makeatother

\usepackage{ragged2e}

\usepackage{parallel,enumitem}

\newtheorem{definition}{Definition}

\algnewcommand{\Inputs}[1]{%
  \State \textbf{Inputs:}
  \Statex \hspace*{\algorithmicindent}\parbox[t]{.8\linewidth}{\raggedright #1}
}
\algnewcommand{\Outputs}[1]{%
  \State \textbf{Outputs:}
  \Statex \hspace*{\algorithmicindent}\parbox[t]{.8\linewidth}{\raggedright #1}
}
\algnewcommand{\Initialize}[1]{%
  \State \textbf{Initialize:}
  \Statex \hspace*{\algorithmicindent}\parbox[t]{.8\linewidth}{\raggedright #1}
}

\newcommand{\nwu}[1]{{\color{purple}{#1}}}

\begin{document}

\title{zkUnlearner: A Zero-Knowledge Framework for Verifiable Unlearning \\ with Multi-Granularity and Forgery-Resistance}

\author{
{\rm Nan Wang}\Mark{1}, ~{\rm Nan Wu}\Mark{1}\thanks{The corresponding author.}, ~{\rm Xiangyu Hui}\Mark{2}, ~{\rm Jiafan Wang}\Mark{1}, ~{\rm Xin Yuan}\Mark{1}\\
CSIRO's Data61\Mark{1}, Australia, ~The University of Melbourne\Mark{2}, Australia
}

\maketitle

\thispagestyle{empty}

\begin{abstract}
As the demand for exercising the "right to be forgotten" grows, the need for verifiable machine unlearning has become increasingly evident to ensure both transparency and accountability. We present {\em zkUnlearner}, the first zero-knowledge framework for verifiable machine unlearning, specifically designed to support {\em multi-granularity} and {\em forgery-resistance}.

First, we propose a general computational model that employs a {\em bit-masking} technique to enable the {\em selectivity} of existing zero-knowledge proofs of training for gradient descent algorithms. This innovation enables not only traditional {\em sample-level} unlearning but also more advanced {\em feature-level} and {\em class-level} unlearning. Our model can be translated to arithmetic circuits, ensuring compatibility with a broad range of zero-knowledge proof systems. Furthermore, our approach overcomes key limitations of existing methods in both efficiency and privacy. Second, forging attacks present a serious threat to the reliability of unlearning. Specifically, in Stochastic Gradient Descent optimization, gradients from unlearned data, or from minibatches containing it, can be forged using alternative data samples or minibatches that exclude it. We propose the first effective strategies to resist state-of-the-art forging attacks. Finally, we benchmark a zkSNARK-based instantiation of our framework and perform comprehensive performance evaluations to validate its practicality.

\end{abstract}

\section{Introduction}

{\bf Machine unlearning} has garnered significant attention due to the increasing demand for exercising the "right to be forgotten". This process involves removing the influence of specific data from trained models, effectively enabling these models to "forget" the data. Machine unlearning is primarily motivated by the following factors \cite{unlearningmotivation}:
\begin{itemize}
    \item{\bf Access Revocation.} This pertains to cases where sensitive or copyrighted data was inadvertently included in the training process. Once access to such data is revoked, it must be removed from the models to ensure it is no longer utilized or leveraged.

    \item{\bf Model Correction.} When models are trained on outdated knowledge, toxic content, unethical material or misinformation, unlearning offers a mechanism to correct them, ensuring they are no longer influenced by such harmful data.
\end{itemize}

\noindent
{\bf Verifiable unlearning.} Responsible Artificial Intelligence (AI) is a critical and rising trend focused on developing and deploying AI systems that are safe, ethical, and trustworthy, prioritizing fairness and societal well-being. Verifiable unlearning is an essential component of Responsible AI. It empowers data owners to control the use of their personal data, minimizing risks of misuse and unauthorized exploitation, fostering trust with model trainers. Simultaneously, it provides model trainers with a transparent and auditable approach to model correction, enabling compliance with regulatory and ethical standards and building trust with regulators. 

\smallskip
\noindent
{\bf Privacy} is a key consideration in verifiable unlearning. Both model trainers and data owners are driven to safeguard their sensitive information, ensuring that no private information is exposed to unauthorized parties during the unlearning process. Any such exposure could grant adversaries critical insights into the underlying models or datasets, creating serious security vulnerabilities. For example, model trainers aim to protect the privacy of their models, while data owners seek to ensure their datasets are not disclosed to other owners.

\subsection{Prior Works and Limitations}

Although various machine unlearning techniques~\cite{SISA,gupta2021adaptive,Guo2019CertifiedDR,wu2020deltagrad,thudi2022unrolling,chen2021machine,xu2023survey,Thudi2021OnTN} have been studied, verifiable unlearning remains an underexplored field, with only a handful of studies addressing this challenge:

\smallskip
\noindent
{\bf Prediction Comparison.} Comparing model predictions offers an intriguing avenue for exploring verifiable unlearning. This line of research works \cite{sommer2022athena, gao2022, backdoorunlearning} achieves unlearning by injecting backdoors into the unlearned data and verifying the success of unlearning by comparing the model's predictions before and after the process. However, recent research finds this approach less reliable. Shumailov et al. (NeurIPS \textquotesingle21) \cite{dataorderingattack} demonstrates that data ordering can introduce {\em randomness} into model training, showing that using different datasets can produce similar or even identical model updates, resulting in models with undetectable differences in predictions. Based on this finding, a recent work (USENIX Sec \textquotesingle22) \cite{Thudi2021OnTN} presents a novel {\em forging attack}, allowing attackers to use a dataset containing unlearned data to train a model that mimics the expected behavior of a model trained using a different dataset excluding those data. Thus, it becomes challenging to determine whether specific data has been unlearned solely by comparing model parameters or predictions. Moreover, the study redefines verifiable unlearning at the algorithmic level rather than the parameter level, emphasizing verifiable computation as a promising solution for ensuring reliable unlearning.

\smallskip
\noindent
{\bf Verifiable Computation.} Verifiable computation is considered an ideal solution for defending against forging attacks in verifiable unlearning, as it can provably replicate prescribed computations tied to model training on designated datasets. However, it does not inherently prevent such attacks. Particularly, in model training with Minibatch Stochastic Gradient Descent (MSGD) optimization, where model updates are based on randomly selected minibatches from training datasets, a malicious model trainer can strategically select a {\em forging minibatch} excluding unlearned data to generate model updates that closely resemble those produced by a target minibatch containing it. As a result, training the model with the forging minibatch is equivalent to training it with the target minibatch, thereby preserving the influence of the unlearned data. Two recent studies \cite{trustedhardware, eisenhofer2023} utilize different verifiable computation approaches, namely, Trusted Execution Environments (TEEs) and Zero-Knowledge Proofs (ZKPs), to achieve verifiable unlearning, respectively. However, neither of them resolves the fundamental issue of resisting forging attacks.

\subsection{Research Gaps}

\noindent
To date, two significant research gaps remain unaddressed:

\smallskip
\noindent
{\bf Forging Attacks.} To date, forging attacks remain a significant threat to the reliability of unlearning~\cite{LI2024100254,suliman2024data}. For MSGD optimization, the core issue lies in the adversary-controlled randomness, which enables manipulation of minibatches to forge model updates that allow the influence of unlearned data to subtly persist in trained models. The challenge intensifies when a {\em gradient replica} of the unlearned data, e.g., a duplicate, exists within the training dataset, capable of producing highly similar gradient updates. Privacy-preserving settings further exacerbate the situation because data owners are unaware of the contributions made by others. Even if one owner's unlearned data is removed, the presence of its gradient replicas within the dataset allows the influence of the unlearned data to persist after unlearning. 

\smallskip
\noindent
{\bf Multi-Granular Unlearning.} Beyond forging attacks, we uncover a new {\em multi-granular} unlearning demand. Traditional studies have primarily focused on sample-level unlearning, which involves removing individual data samples' impact from trained models~\cite{song2017machine,SISA}. For example, a data owner may request the unlearning of specific images from its dataset. However, this paradigm is increasingly falling short of meeting evolving demands, driven by the growing need for more granular and flexible control over the unlearning process:
\begin{itemize}
    \item{\bf Feature-Level Unlearning.} To address access revocation issues, feature-level unlearning has become indispensable. This paradigm allows for the selective removal of specific input features within data samples, ensuring that only targeted aspects of data owners' information are erased, while preserving the remaining data crucial for training. For example, a data owner may request the unlearning of specific pixels revealing sensitive objects, such as faces, in multiple images while retaining the rest.

    \item{\bf Class-Level Unlearning.} To tackle model correction challenges, class-level unlearning is essential for classification tasks, which involve categorizing data into different classes based on specific features or patterns. This unlearning process enables model trainers to correct outdated, mislabeled, or toxic classes within training datasets, thereby enhancing model accuracy and ensuring compliance with ethical standards~\cite{lin2023erm,chen2023boundary}. For example, in a multi-label classification task, a model was trained on a dataset of images containing guns. The model needs to unlearn the pixels associated with guns and remove the gun labels from its training data.
\end{itemize}

\iffalse

\begin{tcolorbox}[colback=red!5!white,colframe=red!75!black,title=Research Gaps]
To date, these two significant research gaps remain unaddressed, as none of the existing studies have
\begin{itemize}
    \item{\em supported multi-granularity in verifiable unlearning.} 

    \item{\em addressed forging attacks targeting SGD optimization.}
\end{itemize}
\end{tcolorbox}
\fi

\section{Contribution} \label{sec:contribution}

In this paper, we explore the challenge of ZKP-based verifiable unlearning and aim to bridge existing gaps by presenting {\em zkUnlearner}, the first zero-knowledge framework designed for {\em multi-granular} and {\em forgry-resistant} verifiable unlearning. Exact unlearning is a robust solution for addressing critical issues, such as copyright infringement and privacy as it can completely remove the influence of problematic data by retraining models from scratch. Given its effectiveness, we focus on this paradigm in our work.

\begin{figure*}[!ht]
    \centering
    \includegraphics[width=0.64\linewidth,trim={0 1.3cm 0 0},clip]{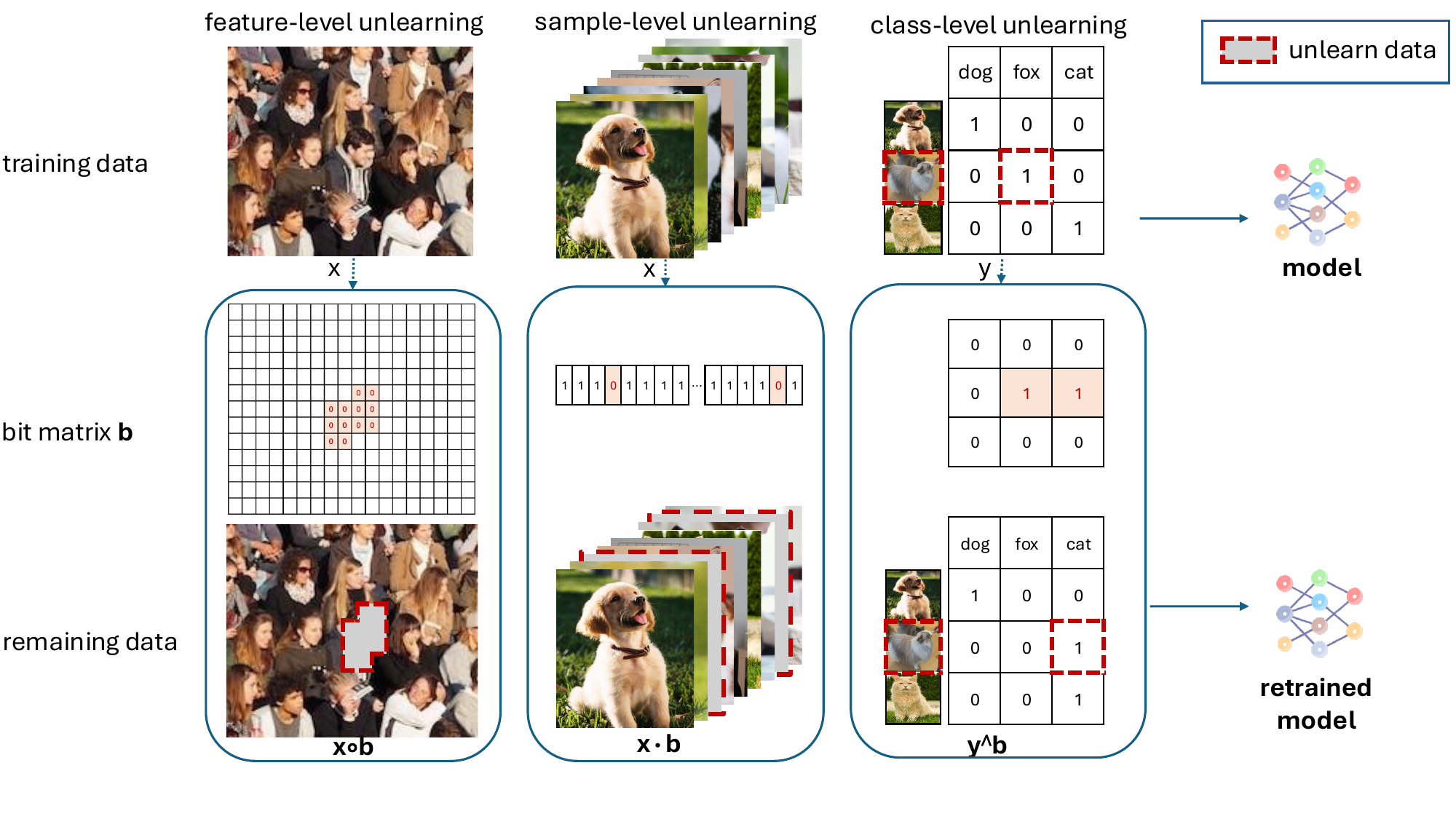}
    \caption{An overview of bit-masking-based multi-granular unlearning, with a one-hot encoding-based class-level unlearning.}
    \label{fig:illustration}
\end{figure*}

\subsection{Verifiable (Un)learning with zkPoTs}

We begin with a brief background of ZKP-based verifiable (un)learning before elaborating on our contributions. ZKPs are a popular cryptographic primitive, allowing a prover to convince a verifier of the validity of a statement without revealing any other information. Using zero-knowledge proofs of training (zkPoT) has proven to be a robust method for achieving verifiable machine learning (CCS \textquotesingle24, \textquotesingle23)\cite{PoT1, PoT2} in a privacy-preserving manner. A zkPoT enables a model trainer to demonstrate the correct training of a committed model on a committed dataset without disclosing any information about the dataset or the model. Specifically, a zkPoT enables verifiable gradient descent computations by modeling them as arithmetic circuits, a standard computational model for computing polynomials that forms the basis for constructing general-purpose zero-knowledge proofs, such as zkSNARKs. A zkPoT can also facilitate verifiable unlearning \cite{eisenhofer2023}. Upon receiving an unlearning request, the model trainer retrains the model using the updated dataset excluding unlearned data, and generates a new zkPoT to validate the process. 

\subsection{Multi-Granular Unlearning}

{\bf {\em We devise a general computational model that employs a bit-masking technique to enable the selectivity of existing zkPoTs, providing flexible control of specific data units' impact on trained models so as to achieve multi-granular unlearning.}} Additionally, our technique preserves the features of these zkPoTs.

\begin{figure}[!h]
	\centering
	\includegraphics[width=0.45\textwidth]{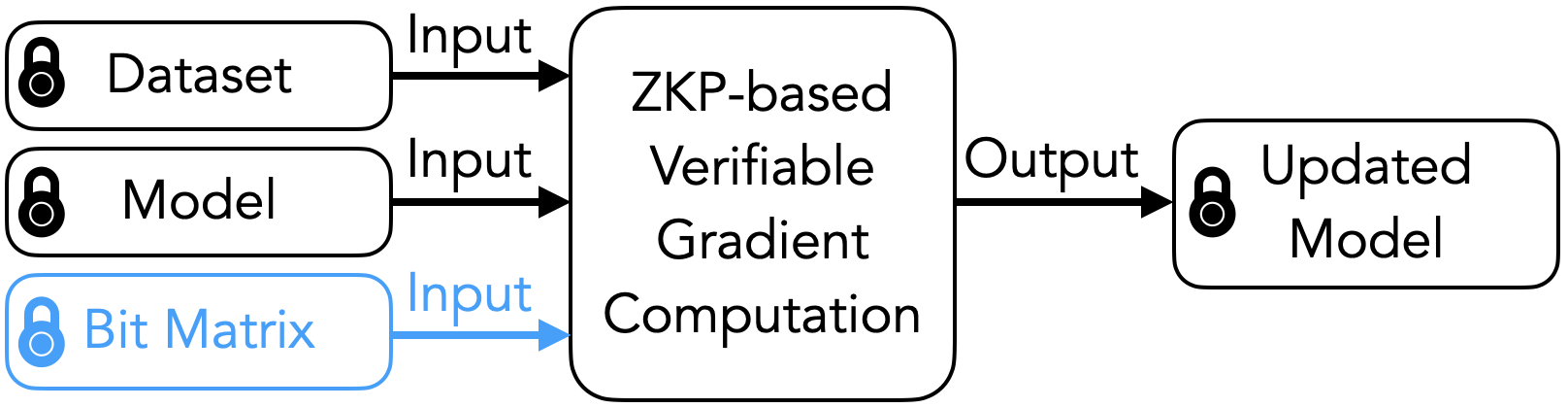}
	\caption{Our Computational Model.}
	\label{fig:vc}
\end{figure}

\noindent
{\bf Intuition.} As shown in Figure \ref{fig:vc}, our bit-masking technique innovatively incorporates an additional committed {\em unlearning bit matrix} into the construction of existing zkPoTs. {\bf {\em At a high level, our technique enables the replacement of data units, such as features, class labels, and samples, with their bit-masked counterparts in all gradient descent computations. This guarantees both the removal and correction of the influence of specific data units on trained models, driven by the goals of access revocation and model correction, while preserving privacy.}} Each bit controls a data unit by applying a specific bit-wise operation, such as {\em AND}, {\em OR} or {\em Exclusive-OR}. Notably, our approach is universally applicable to all gradient descent-based machine learning models, including neural networks, linear regression, and other models with differentiable kernels, as these models rely on gradient descent algorithms as the foundation of their optimization processes. Figure \ref{fig:illustration} offers a high-level overview of how our bit-masking technique facilitates verifiable unlearning across multiple granularities.

\smallskip
\noindent
{\bf Advantages.} Our technique overcomes several key limitations of the state-of-the-art ZKP-based unlearning framework \cite{eisenhofer2023}. These include support for multi-granular unlearning, enhanced privacy by preventing the leakage of indices or the quantity of unlearned data, and invariable arithmetic circuits and datasets for significantly improved efficiency. A detailed overview of the advantages is provided in Section \ref{sec:limitation}.

\begin{figure}[!h]
	\centering
	\includegraphics[width=0.45\textwidth]{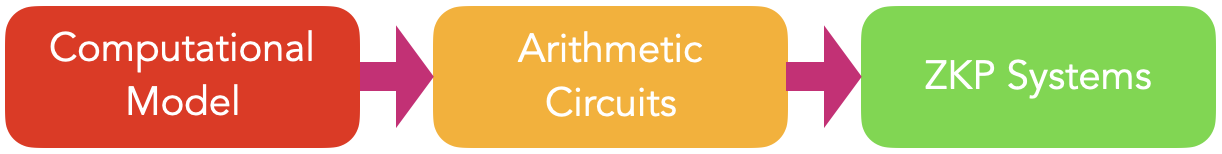}
	\caption{A general workflow of ZKP constructions.}
	\label{fig:model}
\end{figure}

\smallskip
\noindent
{\bf Instantiation.} As illustrated in Figure \ref{fig:model}, our model can be converted into arithmetic circuits, ensuring compatibility with a variety of zero-knowledge proof systems for practical implementation. We implement our framework using a popular zkSNARK, Groth16 \cite{groth16}, and evaluate the performance to demonstrate its practicality.

\smallskip
\noindent
{\bf Extensions.} Throughout the paper, we focus on the centralized environments, where model trainers have access to data owners' datasets and use them to train models. Our technique can be further extended to support the SISA framework and federated environments.

\subsection{Resistance To Forging Attacks}

\noindent
Existing studies \cite{Thudi2021OnTN, zhang24h} have presented two different approaches to forging model updates in SGD-optimized training, and we propose effective solutions to resist them:

\smallskip
\noindent
{\bf Adversary-Controlled Random Sampling.} Model trainers may exploit random sampling to strategically assemble specific data samples into mini-batches during retraining. To mitigate this vulnerability, {\bf {\em we propose the first solution that introduces publicly verifiable randomness into SGD-optimized training, ensuring that the randomness cannot be manipulated by adversaries.}} Our solution drastically reduces the likelihood of successfully executing such attacks. We utilize the well-established cryptographic primitive, {\em Verifiable Random Functions (VRFs)} \cite{VRF}, to securely and provably select random data samples or minibatches from training datasets.

\smallskip
\noindent
{\bf Replacement with Closest Class-Wise Neighbor.} This approach replaces the unlearned data in a minibatch with its closest class-wise neighbor in the dataset. Specifically, we found out that the attack becomes highly effective when this neighbor happens to be a gradient replica. Our publicly verifiable randomness cannot prevent this issue, as any minibatch containing such gradient replicas can produce forged gradients regardless of randomness. Thus, we provide an approach to detect such potential forging risks: {\bf {\em we require model trainers to produce a zero-knowledge proof demonstrating the similarity between the model updates generated by each data sample in the selected minibatch and the unlearned data. Data samples with a similarity below a specified threshold are flagged as suspicious, as they may facilitate forging attacks.}} We will discuss possible strategies for resisting such attacks once they are detected.

\subsection{Roadmap}

We formulate the problem of ZKP-based verifiable unlearning in Section \ref{sec:problem}. We present preliminaries of machine learning concepts and cryptographic tools in Section \ref{sec:machinelearning} and \ref{sec:cryptography}, respectively. We review the limitations of the existing approach and describe our techniques in Section~\ref{sec:limitation} and Section~\ref{sec:approach}, respectively. We describe our approach of resisting forging attacks in Section~\ref{sec:forgingattack}. We briefly discuss some extensions and evaluate an instantiation of our framework in Section~\ref{sec:extension} and Section~\ref{sec:experiment}.

\section{Problem Formulation} \label{sec:problem}

In this section, we define a general centralized machine unlearning problem involving a group of {\em data owners}, each possessing a dataset, and a {\em model trainer} responsible for training machine learning models using the datasets provided by the data owners.

\smallskip
\noindent
{\bf Threat Model.} \nwu{We consider a realistic setting involving both honest-but-honest entities and potential malicious entities.} We make a practical assumption: the model trainer is committed to training high-quality models and has no incentive to compromise their performance. However, the trainer may be tempted to cheat and retain the influence of unlearned data if such data positively \nwu{contributes} the model's performance. Conversely, data owners might have motives to sabotage the model, prompting the model trainer to take measures to prevent such malicious actions. 
\nwu{So, the unlearning process must be auditable to prevent such non-compliance.}
\nwu{On the other hand, data owners may act maliciously and potentially submit modified data or trigger unlearning requests with the intention of degrading model quality. Because of the two-way threat between the model trainer and data owners, both contributions and deletions are required to be secure and verifiable.}
Additionally, we assume the presence of a peer-to-peer secure communication channel between the model trainer and each data owner, ensuring the secure exchange of messages. In the centralized setting, it is inherently impossible to prevent the model trainer from covertly leaking data owners' datasets to third parties, given its direct access to the data. However, the model \nwu{trainer} is unlikely to engage in such behavior, as data serves as the cornerstone of models, helping to establish technical barriers and maintain a competitive edge in real-world scenarios. Therefore, our framework enforces strict access control at the protocol level, effectively prohibiting unauthorized access. For instance, during the unlearning process, each data owner's dataset must remain confidential from other data owners, and the trained models must also remain private.

\smallskip
\noindent
{\bf Verifiable Unlearning.} We describe a general ZKP-based exact unlearning scenario. When a data owner submits an unlearning request for a committed model trained on datasets from multiple owners, she privately provides the model trainer with the indices of the data to be unlearned from her committed dataset. Then the model trainer retrains the model using the updated dataset excluding the specified unlearned data and publishes a commitment to the resulting updated model. Finally, the model trainer provides a new zkPoT to prove that the updated committed model was trained using the updated committed datasets. Note that data owners can submit unlearning requests at any time, and a single unlearning process can accommodate multiple data owners' requests simultaneously.

\smallskip
\noindent
{\bf Challenge.} The core challenge lies in developing an efficient way of provably ensuring that the updated dataset is indeed derived from the original training dataset by fully and accurately removing the specified unlearned data while preserving privacy. Importantly, the solution must be robust against malicious model trainers that attempt to covertly add new unauthorized data to the training dataset or remove unspecified data from the training dataset. Such actions could facilitate forging attacks, as malicious model trainers may manipulate the data ordering within minibatches to forge model updates.

\section{Machine Learning Concepts} \label{sec:machinelearning}

Machine learning involves training a model to learn patterns and make predictions based on a given dataset. There are two major types: {\em supervised} and {\em unsupervised} learning. The key distinction lies in the nature of the dataset: supervised learning operates on a labeled dataset, while unsupervised learning deals with unlabeled data. This paper focuses specifically on supervised learning. A supervised learning model is trained on a dataset ${\bf D}$ consisting of data samples $({\bf x}_i, y_i)_{i=1}^N$, where ${\bf x}_i$ and $y_i$ represent the input feature vector and corresponding labeled output of the $i$-th sample. The goal aims to learn from ${\bf D}$ and accurately predict labels $\hat{y}_i$ for new input data. Note that we denote vectors and matrices using bold notation throughout the paper.

\subsection{Loss Function}

The loss function, also known as the cost function, measures how well a machine learning model's predictions match the actual data. It quantifies the difference between the predicted values and the true values. Loss functions act as the guiding metric for optimization algorithms, such as gradient descent, that iteratively adjust model parameters to minimize errors and improve performance.

\subsubsection{Regression tasks}

Regression tasks involves predicting continuous numerical values. A commonly used loss function for regression tasks is the Mean Squared Error (MSE):
    \[
    {\mathcal{L}}({\bf w}) = \frac{1}{N} \sum_{i=1}^{N} (y_i - \hat{y}_i)^2
    \]
    where \( y_i \) and \( \hat{y}_i \) are the observed and predicted values, \( \boldsymbol{w} \) are the model parameters, and \( N \) is the number of data samples. 
    
\subsubsection{Classification tasks}

Classification tasks assign data to discrete categories. Cross-Entropy Loss is employed for classification tasks: 
 \[
    {\mathcal{L}}({\bf w}) = -\frac{1}{N} \sum_{i=1}^{N} \left[ y_i \log(\hat{y}_i) + (1 - y_i) \log(1 - \hat{y}_i) \right]
    \]
where \( y_i \) and \( \hat{y}_i \) are the observed label and the predicted probability, respectively. 

\smallskip
\noindent
{\bf Subtypes.} There are two types of classification tasks: 
\begin{itemize}
   \item{\bf Multi-class classification} assigns each data sample to one mutually exclusive output class, such as categorizing an image as a dog, cat, or bird.

    \item{\bf Multi-label classification} extends multi-class classification by allowing a data sample to belong to multiple output classes simultaneously. For example, an image may contain both a dog and a cat, requiring two labels. 
\end{itemize}

\begin{table}[t!]
    \centering
    \caption{Key Symbols and Notations}
    \resizebox{0.45\textwidth}{!}{%
    \begin{tabular}{@{}cc@{}}
    \toprule
    Symbols          & Descriptions  \\ \midrule
    $({\bf x}, y)$ & A data sample (input features, output label) \\
    $({\bf x}_u, y_u)$ & The unlearned data sample \\
    ${\bf D}$ & The training dataset \\ 
    ${\bf U}$ & The list of unlearned data \\ 
    ${\bf d}$ & The minibatch in MSGD-optimized training \\ 
    $T$ & The number of unlearning rounds  \\
    $N$ & The number of data samples in a dataset   \\
    $J$ & The number of input features in a data sample   \\
    $K$ & The number of output class labels in a data sample   \\
    ${\bf w}$ & The parameters of trained models  \\
    ${\mathcal{L}}({\bf w}, {\bf d})$ & The loss function w.r.t {\bf w} and {\bf d}\\ 
    $\nabla {\mathcal{L}}({\bf w}, {\bf d})$ & The gradient w.r.t {\bf w} and {\bf d} \\ 
    $\pi$ & Zero-knowledge proof \\
    $|~|$ & The cardinality of a set \\
    $||~||$ & The Euclidean norm  \\
    $\circ$ & The Hadamard product operator \\
    $\cdot$ & The multiplication operator \\
    $\&$ & The bitwise AND operator \\
    $||$ & The bitwise OR operator \\
    $\hat{~}$ & The bitwise XOR (Exclusive-OR) operator \\
    $\backslash$ & Excluding operator \\
    $\wedge$ & Logical AND operator \\
\bottomrule
\end{tabular}
}
\label{tab:symbols}
\end{table}

\noindent
{\bf Encodings.} In classification tasks, class labels are often represented using various binary encodings, such as multi-hot encoding or ordinal encoding, where each label is expressed as a binary vector. Multi-hot encoding, in particular, allows multiple "1"s within a vector to represent multiple active labels, while "0"s denote inactive labels. It is commonly applied in multi-label classification tasks. This approach generalizes the widely used {\em one-hot encoding}, which limits the vector to a single "1" and is primarily utilized in multi-class classification tasks. For instance, in a multi-class classification problem with three labels \{dog, fox, cat\}, a dog image could be encoded as [1, 0, 0], while a fox image would be represented as [0, 1, 0].

\subsection{Optimization Function}

Optimization functions are the driving force behind machine learning, enabling models to minimize errors and improve performance. Gradients serve as the backbone of optimization algorithms. By indicating the direction and rate of change of a model's loss function with respect to its parameters, gradients guide the iterative adjustment of weights to minimize errors. 
\[
{\bf w}^{(t+1)} = {\bf w}^{(t)} - \eta \cdot \nabla {\mathcal{L}}\big({\bf w}^{(t)}, ({\bf x}, y)\big)
\]
where \( \eta \) is the learning rate, \({\bf w}^{(t)} \) are the parameters at iteration \( t \), and \( \nabla {\mathcal{L}}({\bf w}^{(t)}) \) is the gradient of the loss function with respect to the parameters. There are three optimization functions commonly used during model training:

\smallskip
\noindent
{\bf Batch Gradient Descent (BGD).} This optimization function computes gradients using the entire training dataset in each epoch. This approach is often preferred when working with small datasets or aiming for high stability in convergence.  
\[
{\bf w}^{(t+1)} = {\bf w}^{(t)} - \frac{\eta}{|{\bf D}|}  \cdot 
\sum_{i=1}^{|{\bf D}|} \nabla {\mathcal{L}}\big({\bf w}^{(t)}, ({\bf x}_i, y_i)\big)
\]

\smallskip
\noindent
{\bf Stochastic Gradient Descent (SGD).} This optimization function updates model parameters using a randomly sampled data sample $({\bf x}, y) \in {\bf D}$ at each step. SGD enables faster updates, reduces memory requirements, and introduces a degree of stochasticity that can help escape local minima. 

\smallskip
\noindent
{\bf Minibatch Stochastic Gradient Descent (MSGD).} This optimization function updates model parameters using a randomly sampled minibatch ${\bf d} \subset {\bf D}$ at each step. It is widely used in machine learning tasks due to its efficiency and scalability.
\[
{\bf w}^{(t+1)} = {\bf w}^{(t)} - \frac{\eta}{|{\bf d}|}  \cdot 
\sum_{m=1}^{|{\bf d}|} \nabla {\mathcal{L}}\big({\bf w}^{(t)}, ({\bf x}_m, y_m)\big)
\]
%%In our paper, we consider 
SGD is considered as a special case of MSGD with $|{\bf d}|=1$.

\smallskip
\noindent
{\bf Remark:} {\em epoch} and {\em step} are two important terms that describe the training process, especially in SGD-optimized training. An {\em epoch} refers to a complete pass through the entire training dataset during the training process while a {\em step} refers to a single update of the model's parameters during training, based on one batch of data.

\iffalse
\subsection{Machine Unlearning}

Exact unlearning and approximate unlearning are two paradigms for removing the influence of specific data from a machine learning model. 

\smallskip
\noindent
{\bf Exact unlearning} ensures that the model is completely retrained as if the data to be unlearned had never been included in the training process. This approach is often computationally expensive, as it requires retraining the model from scratch, but it provides strong guarantees of compliance with data removal requests. 

\smallskip
\noindent
{\bf Approximate unlearning} seeks to efficiently approximate the effect of removing data by updating the model in a way that mimics the outcome of exact unlearning. While this approach is computationally cheaper and more scalable, it typically sacrifices some precision, relying on techniques such as gradient updates or model fine-tuning. The trade-off between exactness and efficiency makes the choice between these paradigms application-dependent, often balancing regulatory requirements, resource constraints, and model performance.
\fi

\section{Cryptographic Tools} \label{sec:cryptography}

We provide a concise overview of essential cryptographic tools. We use $\lambda$ to represent the security parameter. Let $\xleftarrow{\$}$ indicate random sampling.

\subsection{Commitment Scheme}

In this work, we use vector commitment schemes. A vector commitment scheme is a cryptographic primitive that allows one party to commit to a vector of integer values while keeping them hidden from others, with the ability to reveal the committed values later. A vector commitment scheme consists of three probabilistic polynomial-time (PPT) algorithms ({\sf Setup}, {\sf Commit}, {\sf Open}):
    \begin{itemize}
    \item{\bf Setup:} on input a security parameter $\lambda$, generates the public commitment key.
    
    \item{\bf Commit:} on input a message vector ${\bf m}$ and the randomness $r$, outputs a commitment value ${\sf Com}({\bf m};r)$.
    
    \item{\bf Open}: on input a commitment ${\sf Com}({\bf m};r)$, a message vector ${\bf m}$ and the randomness $r$, outputs a bit indicating the validity of the commitment.
\end{itemize}
A commitment scheme must be: 1) {\em hiding}, such that the commitment does not reveal information about the committed values; 2) {\em binding}, such that a commitment can only be opened to a unique message vector.  

\subsection{Hash Function}

Hash functions are fundamental cryptographic primitives widely used in security applications. A hash function ${\mathcal{H}}(\cdot): \{0, 1\}^* \rightarrow \{0, 1\}^{fix}$ takes an input of arbitrary size and produces a fixed-size output, often called a hash or digest. A hash function must satisfy: 1) {\em one-wayness}, such that the same input always produces the same output; 2) {\em collision-resistance}, such that it is computationally infeasible to find two different inputs with the same output; 3) {\em preimage-resistance}, such that it is computationally hard to deduce the input from its output. Note that hash functions can be seen as instances of commitment schemes.

\subsection{Zero-Knowledge Proofs of Knowledge} \label{sec:zkp}

A zero-knowledge proof of knowledge (ZKPoK) is an interactive protocol for relation $R$ with a language $L$ between a prover ${\mathcal{P}}$ and a verifier ${\mathcal{V}}$ on a common statement $u$. The protocol must satisfy: 1) {\em completeness}, such that ${\mathcal{P}}$ holding a witness $\omega$ can always convince ${\mathcal{V}}$ of $(u, \omega) \in R$; 2) {\em soundness}, such that ${\mathcal{P}}$ cannot convince ${\mathcal{V}}$ to accept when $u \notin L$ with overwhelming probability; 3) {\em zero-knowledge}, such that ${\mathcal{V}}$ learns no information about $\omega$ other than $(u, \omega) \in R$.

\iffalse

\begin{definition}[Zero-Knowledge Proof of Knowledge]
A zero-knowledge proof of knowledge for a language ${\mathcal{L}}$ is an interactive protocol between a prover ${\mathcal{P}}$ and a verifier ${\mathcal{V}}$. ${\mathcal{P}}$ knows a witness $\omega$ such that the language ${\mathcal{L}}$ holds for $\omega$ and $u$. A ZKPoK consists of a triple of PPT algorithms ({\sf Setup}, {\sf Prove}, {\sf Verify}): 
    \begin{itemize}
    \item{\bf Setup:} ${\sf pp} \leftarrow {\sf Setup}(\lambda)$ generates the public parameters.
    
    \item{\bf Prove:} $\pi \leftarrow {\sf Prove}_{\sf pp}(u, \omega, {\mathcal{L}})$ takes as inputs a statement $u$ and a witness $\omega$ and a language ${\mathcal{L}}$. It outputs a proof $\pi$.
    
    \item{\bf Verify}: $b \leftarrow {\sf Verify}_{\sf pp}\big(u, \pi\big)$ takes as inputs a statement $u$, a proof $\pi$. It outputs a bit indicating whether the proof is valid.
    \end{itemize}
\end{definition}
A ZKPoP must satisfy the three key security properties:
\begin{itemize}
    \item{\bf Correctness:} ${\mathcal{P}}$ can always convince ${\mathcal{V}}$ if the proof is valid.
    
    \item{\bf Soundness:} ${\mathcal{P}}$ cannot convince ${\mathcal{V}}$ if the proof is not valid.

    \item{\bf Zero-Knowledge:} ${\mathcal{V}}$ learns nothing except the validity of the proof.

\end{itemize}
\fi

\begin{figure}[!h]
	\centering
	\includegraphics[width=0.25\textwidth]{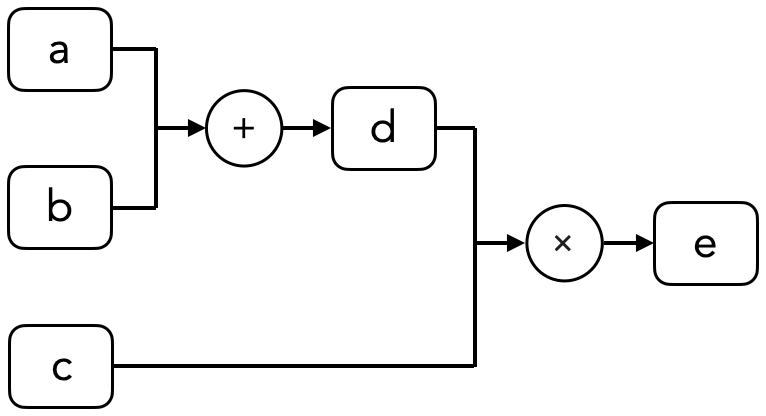}
	\caption{An example of an arithmetic circuit.}
	\label{fig:arithmeticcircuit}
\end{figure}

\smallskip
\noindent
{\bf Arithmetic circuits} are a low-level representation of a program that consists of inputs, gates, and outputs, where each gate performs basic operations of addition and multiplication over values from a finite field. Figure \ref{fig:arithmeticcircuit} illustrates an arithmetic circuit example for computing a polynomial $e=(a+b) \cdot c$, where ($a$, $b$, $c$) are the inputs, $e$ is the output and $d$ is an intermediate output. {\bf {\em Note that if the structure of the computation changes,  then the arithmetic circuit must be updated to reflect these changes.}} For instance, removing the input from the circuit will alter the structure.

\smallskip
\noindent
{\bf Zero-Knowledge Succinct Non-interactive Arguments of Knowledge (zkSNARKs)} are a specific type of zero-knowledge proof systems with short proof sizes and can be verified in time sublinear in the size of the statement. The constraints of zero-knowledge proofs can be translated into arithmetic circuits for verifiable computation. zkSNARKs implement arithmetic circuits by transforming the computational problem into a series of polynomial equations, enabling the prover to demonstrate that they have a valid input-output pair that satisfies the circuit, without revealing any details about the inputs or outputs. 

\subsection{Verifiable Random Functions}

Verifiable random functions (VRFs) \cite{VRF} are cryptographic primitives that provide a way to generate a pseudorandom output and a proof of its correctness, which can be publicly verified. A VRF scheme consists of three PPT algorithms:
\begin{itemize}
    \item{\bf Key Generation:} on input a security parameter $\lambda$, generates a secret key {\sf sk} and a public key {\sf pk}.

    \item{\bf Evaluation:} on input a string $\mu$ and a secret key {\sf sk}, generates a pseudorandom output $r$ and a proof $\pi$ showing that $r$ was correctly computed.
    
    \item{\bf Verification:} on input $r$, $\mu$, $\pi$ and a public key {\sf pk}, generates a bit $b$ indicating the validity of the proof. 
\end{itemize}
The key generation phase utilizes a public-key infrastructure (PKI) to generate a key pair for each user. Within our framework, the model trainer registers with the PKI and obtains its own key pair to use VRFs.

\section{Overview of Existing ZKP-Based Approach} \label{sec:limitation}

In this section, we provide an in-depth analysis of the ZKP-based verifiable unlearning framework \cite{eisenhofer2023}, which is the only study directly related to our research. The framework proposes an idea of using non-membership zero-knowledge proofs to support sample-level verifiable unlearning, where a membership proof aims to demonstrate a committed value belongs to a public list without revealing the value.

\subsection{Intuition}
In their framework, the model trainer manages two ordered lists: one for the training data samples provided by data owners, denoted as {\bf D}, and another for the unlearned data samples, denoted as {\bf U}. Initially, {\bf D} contains hash values of the data samples rather than their plain values to safeguard the privacy of the data owners' datasets, while {\bf U} is empty. The model trainer also computes root hash values for both lists, where each root is a recursive hash of all elements in the respective list, serving as a commitment.

When a data owner submits an unlearning request, the model trainer transfers the hash values of the unlearned data samples from {\bf D} to {\bf U}, effectively updating the training list to ${\bf D'}={\bf D} \backslash {\bf U}$. Subsequently, the model trainer retrains the model using the updated dataset ${\bf D'}$. To ensure transparency and accountability, the model trainer generates a zkPoT that establishes verifiable links between the retrained model and the updated dataset ${\bf D'}$, proving that the designated data has been correctly unlearned. 

Additionally, the model trainer must provide a membership proof to prove that the unlearned data samples are included in {\bf U}. This proof consists of a linear-sized hash-chain path from the specific unlearned data sample to the root hash of {\bf U}. According to their reasoning, proving membership in {\bf U} inherently implies non-membership in ${\bf D'}$. This design aims to optimize both communication and computational efficiency. Since the unlearning list {\bf U} is significantly smaller than the training list {\bf D}, generating a membership proof for {\bf U} requires far less overhead than producing a non-membership proof for ${\bf D'}$. The verifier can independently recompute the root hash of {\bf U} using the provided hash-chain path and compare it to the actual root to ensure the validity of the membership proof. 

\subsection{Limitations}

We identify six major limitations of the existing framework:

\smallskip
\noindent
{\bf Restriction to Sample-Level Unlearning.} Feature-level and class-level unlearning necessitate the removal of specific features and class labels within data samples. During retraining, these removed data units cannot simply be left empty and must be replaced with placeholder values, such as zeros. This limitation arises from the inherent nature of membership proofs, which are designed to determine whether a data unit belongs to a predefined set. These proofs are not designed to handle modifications or dynamic changes within individual samples. On the one hand, the inclusion of placeholder values introduces risks of forging attacks, as these values qualify as unauthorized data. Furthermore, it complicates the distinction between placeholder values and actual training data. On the other hand, using zero placeholder values results in identical hash values, potentially causing privacy leaks. 

\smallskip
\noindent
{\bf Variable Datasets.} Their approach relies on dynamically transferring unlearned data from the training list to the unlearning list. This design poses two significant challenges. On the one hand, the constantly changing unlearning list requires both the model trainers and data providers to repeatedly recompute the roots of the training and unlearning lists. When these lists contain a large number of data samples, the root recomputation becomes highly resource-intensive, leading to significant computational overhead. On the other hand, the dynamic updates make it difficult to ensure that no unauthorized or covertly added data is introduced into the training list. Including such new data can introduce randomness into the process, thereby increasing the risk of forging attacks and undermining the reliability of the unlearning procedure.

\smallskip
\noindent
{\bf Variable Arithmetic Circuits.} Using membership-based proofs fails to preserve underlying arithmetic circuit structures, as removing data from training datasets alters the circuit's inputs and their dependencies. This variability necessitates regenerating the arithmetic circuits per unlearning operation, which is computationally expensive and time-consuming, particularly for complex models or datasets. Additionally, for zero-knowledge proof systems that rely on cumbersome trusted setups, each regeneration of underlying circuits requires a new setup process, significantly increasing overhead and reducing scalability.

\smallskip
\noindent
{\bf Lack of Protection Against Forging Attacks with SGD-optimization.} This framework only works with full-batch gradient descent optimization, where gradients for the original training and retraining are computed over deterministic datasets, respectively, without introducing randomness. It offers no solutions to defend against attacks that exploit SGD-optimized training. 

\smallskip
\noindent
{\bf Privacy Leakage.} Their approach exposes information about the unlearned data, even though each data sample is masked by a hash value. When a data owner requests unlearning, the hash values of the unlearned data are publicly transferred to the unlearning list, revealing the indices and the quantity of unlearned data samples to all parties.

\smallskip
\noindent
{\bf Large Overheads.} Their solution relies on a hash-chain-based membership proof, which identifies a path from the hash of specific data to the root of the unlearning list. This proof consists of a linear-sized sequence of hash values forming the path. When the list becomes larger, the communication and computational overhead become prohibitively high.

\section{Our Bit-Masking Approach} \label{sec:approach}

In this section, we begin by describing how our approach achieves verifiable unlearning with BGD optimization across feature, sample, and class levels. We also emphasize how it addresses the shortcomings of the membership-proof-based approach. In Section \ref{sec:forgingattack}, we will demonstrate its compatibility with SGD optimization. The motivations of {\em access revocation} and {\em model correction} give rise to two unlearning paradigms:
\begin{itemize}
    \item{\bf Removal-based unlearning} includes feature-level and sample-level unlearning, where specific features or entire data samples are excluded from gradient computations to eliminate their influence on the model.

    \item{\bf Correction-based unlearning} involves class-level unlearning, where incorrect class labels are rectified to ensure model accuracy and compliance with regulations.
\end{itemize}

\iffalse

Guided by the principle that \textit{"simplicity is the ultimate sophistication"}, the bit-masking approach is designed to be straightforward, flexible, and adaptable to diverse unlearning scenarios. The figure illustrates three examples of deleting or modifying training data and labels. While a simple binary scheme—using bit 1 to retain and bit 0 to delete via logical AND—is often sufficient, the method naturally extends to more sophisticated edits. By encoding the target information as a binary vector and leveraging bitwise XOR operations, data or labels can be selectively updated, as demonstrated in the class-level unlearning scenario.

\fi

\subsection{Feature-Level Unlearning} \label{sec:featurelevelunlearning}

In this section, we exemplify the application of a generalized version of our bit-masking technique to feature-level unlearning scenario. With minor adaptations, the technique can be extended to support unlearning at the sample level (Section~\ref{sec:sampleunlearning}) and the class level (Section~\ref{sec:classunlearning}), respectively.

\subsubsection{Overview}

Our bit-masking technique allows a prover to incorporate a committed {\em unlearning bit matrix} into the verifiable gradient computation process without disclosing any information about the matrix itself. By using bit-wise AND operations, this matrix facilitates the selective removal of specific features' impact on the trained models. Given an $N \times J$ input feature matrix ${\bf x}$ of a dataset, a bit matrix {\bf b} of the same dimension is required, where each bit $b_{i, j}$ is either 1 or 0:
$$
{\bf x}=\begin{pmatrix}
x_{1, 1} & \hdots & x_{1, J} \\
\vdots  & \ddots & \vdots \\
x_{N, 1}  & \hdots & x_{N, J} \\ 
\end{pmatrix} \quad 
{\bf b}=\begin{pmatrix}
b_{1, 1} & \hdots & b_{1, J} \\
\vdots  & \ddots & \vdots \\
b_{N, 1} & \hdots & b_{N, J} \\ 
\end{pmatrix}
$$
{\bf {\em The trick is to replace the input feature matrix ${\bf x}$ with the bit-masked feature matrix ${\bf x} \circ {\bf b}$ in all gradient calculations}}, where $\circ$ denotes the Hadamard product operation. Note that this bit-masked matrix ${\bf x} \circ {\bf b}$ represents the updated dataset. When $b_{i, j}=0$, the impact of the input feature $x_{i, j}$ on the gradients is nullified. In contrast, when $b_{i, j}=1$, the influence of the input feature $x_{i, j}=0$ remains intact.
$$
{\bf x} \circ {\bf b}=\begin{pmatrix}
x_{1, 1} \cdot b_{1, 1} & \hdots & x_{1, J} \cdot b_{1, J} \\
\vdots  & \ddots & \vdots \\
x_{N, 1} \cdot b_{N, 1} & \hdots & x_{N, J} \cdot b_{N, J} \\ 
\end{pmatrix} 
$$

\noindent
{\bf Remark:} When data owners submit unlearning requests, they privately reveal the bit matrix to the model trainer. Then the model trainer publishes a commitment to the bit matrix for each data provider. For those data owners who do not submit unlearning, the bit matrix is the default matrix of one. Each data owner subsequently checks and signs the commitment as confirmation. Finally, the model trainer retrains an updated model using the bit-masked feature matrices from the data owners and provides a zkPoT to demonstrate that the updated model was trained using the updated committed datasets.

\subsubsection{Multi-Variate Linear Regression}

To ease the understanding of our bit masking technique, we exemplify its application to the multi-variate linear regression model $y = \sum_{j=1}^J x_j w_j + \delta$, where $(x_j)_{j=1}^J$, $(w_j)_{j=1}^J$ and $\delta$ represent the input features, the weights and bias, respectively. By applying our bit-masking technique, the gradients are then reformulated from Eqn. \raf{eqn:gradient} to Eqn. \raf{eqn:updatedgradient}:
\begin{align}
\begin{split}
 \nabla {\mathcal{L}} \big((w_j)_{j=1}^J\big) = \frac{1}{N} \sum_{i=1}^N x_{i, j} \cdot (\sum_{j=1}^J x_{i,j} \cdot  w_j + \delta - \hat{y}_i)  
 \label{eqn:gradient}
\end{split} \\
\begin{split}
 \Longrightarrow~\frac{1}{\hat{N}} \sum_{i=1}^N (x_{i, j} b_{i, j}) \cdot \big(\sum_{j=1}^J (x_{i, j} b_{i, j}) \cdot w_j + \delta - \hat{y}_i\big)  
 \label{eqn:updatedgradient}
\end{split} 
\end{align}
where $y$ and $\hat{y}$ are the observed and predicted values, respectively. $\hat{N}$ refers to the number of data samples excluding those that are unlearned. Calculating $\hat{N}$ involves a slightly intricate process. Specifically, we flatten the bit matrix column-wise to obtain a bit vector ${\bf b}=(b_i)_{i=1}^N$, where $b_i$ represents the union of $(b_{i, j})_{j=1}^J$, indicating whether the $i$-th data sample is unlearned ($b_i=0$) or retained ($b_i=1$):
$$\hat{N}=\sum_{i=1}^N b_i,\quad b_i= b_{i, 1}|| b_{i, 2}...||~b_{i, J}$$
where `||' refers to the bitwise {\em OR} operation.

\smallskip
\noindent
{\bf Remark:} We stress that our technique universally applies to all gradient-based machine learning models as it suffices to replace the input features with the bit-masked input features in the gradient computations. 

\subsubsection{Irrecoverable Unlearning}

\begin{figure}[!ht]
	\centering
	\includegraphics[width=0.45\textwidth]{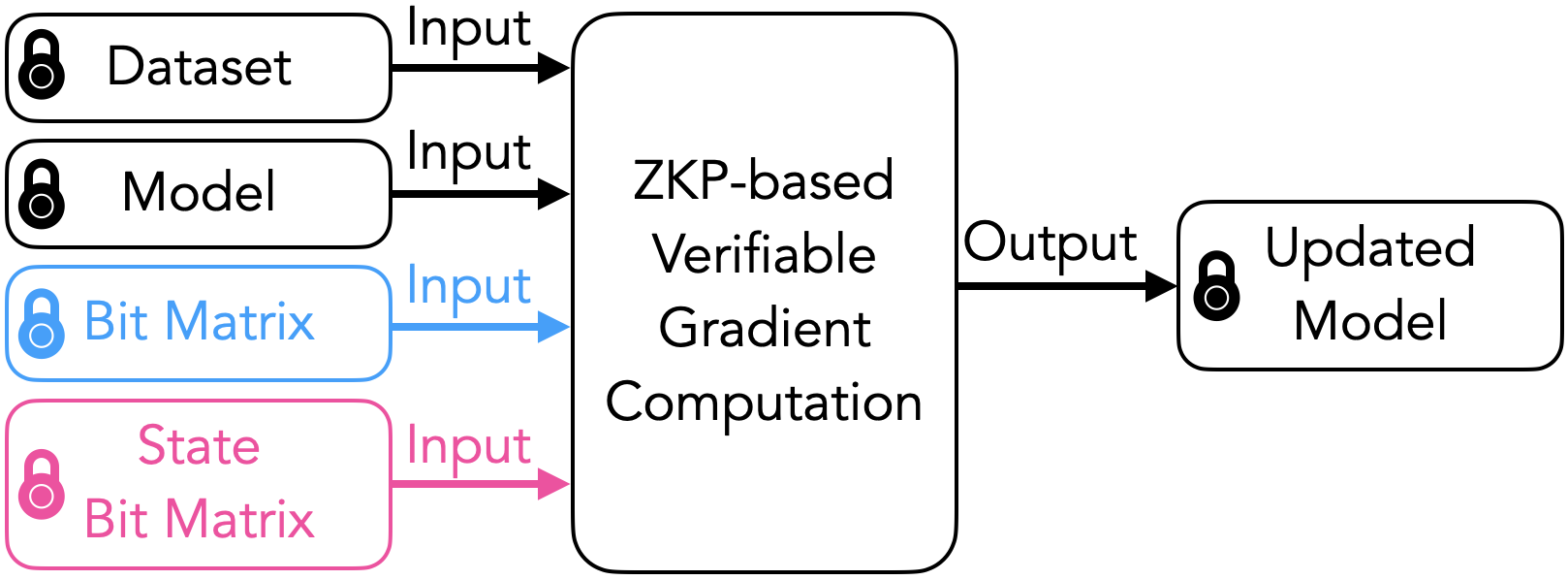}
	\caption{Irrecoverable Verifiable Unlearning.}
	\label{fig:v2}
\end{figure}

To ensure accumulated unlearning and prevent malicious model trainers from reconstructing the unlearned input features across multiple rounds of the unlearning process,  we introduce a novel {\em state-preserving bit matrix} to enhance the unlearning process, as illustrated in Figure \ref{fig:v2}.

In the $t$-th unlearning round, where $t \geq 2 \wedge t \in \{1, ..., T\}$, the prover must utilize the bit matrix ${\bf b}_{t-1}$ from the $(t-1)$-th round as the {\em state bit matrix}. The new bit matrix is generated by combining the current and previous matrices as follows:
$$
{\bf b}_t^{\sf new} \Leftarrow {\bf b}_{t-1}~\&~{\bf b}_t 
$$
where `$\&$' refers to the bit-wise AND operator.

Here, the zero-bits in ${\bf b}_{t-1}$ will persistently nullify the one-bits in ${\bf b}_t$, thereby preserving the unlearning state. For example, as illustrated in Table \ref{tab:recover}, even if a malicious prover attempts to recover the 3rd, 5th, 7th, and 8th bits in the $t$-th round, these bits will remain nullified due to the influence of the preceding bit matrix ${\bf b}_{t-1}$. Notably, in addition to preventing the recovery of unlearned data, our strategy also supports the accumulation of feature unlearning across multiple rounds. For instance, the 6th bit is nullified due to the bit matrix ${\bf b}_t$.

\begin{table}[!ht]
\centering
\caption{An example of the bit-matrix update for irrecoverable unlearning.}
\resizebox{0.42\textwidth}{!}{%
\begin{tabular}{|c|c|c|c|c|c|c|c|c|}
\hline
 & 1st & 2nd & 3rd & 4th & 5th & 6th & 7th & 8th \\ \hline
${\bf b}_{t-1}$ & 1 & 1 & 0 & 1 & 0 & 1 & 0 & 0 \\ \hline
${\bf b}_t$ & 1 & 1 & 1 & 1 & 1 & 0 & 1 & 1 \\ \hline
${\bf b}_t^{\sf new}$ & 1 & 1 & 0 & 1 & 0 & 0 & 0 & 0 \\ \hline
\end{tabular}
}
\label{tab:recover}
\end{table}

\subsubsection{Sample-Level Unlearning} \label{sec:sampleunlearning}

Sample-level unlearning can be seen as a subproblem of feature-level unlearning, where a dimension-reduced bit vector can be utilized for efficiency improvement. Recall that in a multi-variate regression example, all the gradients involve the error term $(y-\hat{y})$. Thus, we directly employ a far smaller $N \times 1$ bit matrix ${\bf b}=(b_i)_{i=1}^N$ in sample-level unlearning, where each entry is a binary indicator representing whether a sample should be included in gradient calculations. It suffices to replace the error $(y_i-\hat{y}_i)$ with $(y_i-\hat{y}_i) \cdot b_i$ without the need for an entire input feature matrix substitution as in feature-level unlearning. 
\begin{align}
\begin{split}
 \nabla {\mathcal{L}}({w_j}) = &~\frac{1}{\hat{N}} \cdot \sum_{i=1}^N x_{i, j} \cdot \big(b_i \cdot (y_i - \hat{y}_i)\big) , \quad \hat{N}=\sum_{i=1}^N b_i
 \label{eqn:sampleupdatedgradient}
\end{split}
\end{align}

\subsection{Class-Level Unlearning} \label{sec:classunlearning}

Unlike selective feature removal in feature-level unlearning, class-level unlearning focuses on achieving model correction by converting specific class labels to desirable ones. To convert between zero bits and one bits, the bit-wise {\em Exclusive OR} operation is required, as the bit-wise AND operation cannot turn zero bits into one bits, making it unsuitable for this purpose. Given a class matrix ${\bf y}$ of the dataset with $K$ class labels, the model trainer generates an $N \times K$ bit matrix ${\bf b}$, where each row represents a data sample and each column corresponds to a class label.
$$
{\bf y}=\begin{pmatrix}
y_{1, 1}  & \hdots & y_{1, K} \\
\vdots   & \ddots & \vdots \\
y_{N, 1} & \hdots & y_{N, K} \\ 
\end{pmatrix} \quad 
{\bf b}=\begin{pmatrix}
b_{1, 1}  & \hdots & b_{1, K} \\
\vdots  & \ddots & \vdots \\
b_{N, 1} & \hdots & b_{N, K} \\ 
\end{pmatrix}
$$
The model trainer applies a similar trick as feature-level unlearning by replacing the class matrix ${\bf y}$ with the matrix ${\bf y} ~\hat{~}~{\bf b}$ in the gradient computations:
$$
{\bf y} ~\hat{~}~ {\bf b}=\begin{pmatrix}
y_{1, 1} ~\hat{~}~ b_{1, 1} & \hdots & y_{1, K} ~\hat{~}~ b_{1, K} \\
\vdots  & \ddots & \vdots \\
y_{N, 1} ~\hat{~}~ b_{N, 1} & \hdots & y_{N, K} ~\hat{~}~ b_{N, K} \\ 
\end{pmatrix} 
$$
where $\hat{~}$ refers to the bit-wise {\em Exclusive-OR} (XOR) operator.

This unlearning involves addressing both multi-class and multi-label classification tasks. Figure \ref{fig:illustration} illustrates a simple multi-class unlearning with one-hot encoding. Here, we provide a more advanced example showcasing a bit matrix for multi-label unlearning utilizing multi-hot encoding\footnote{Our approach also supports alternative encodings supporting bit-wise operations, such as binary encoding, where class labels are represented by their ordinal numbers in binary format.}. This matrix enables the conversion of outputs with the class label $(0, *, *)$ into the label $(1, *, *)$. Here, $*$ represents a wildcard bit, and ${\bf y}$ and ${\bf y'}$ denote the source and target class matrices, respectively:
$$
{\bf y}=\begin{pmatrix}
0 & 0 & 1 \\
1 & 1 & 0 \\
0 & 1 & 1 \\ 
\end{pmatrix} \quad
{\bf b}=\begin{pmatrix}
1 & 0 & 0 \\
0 & 0 & 0 \\
1 & 0 & 0 \\
\end{pmatrix} \quad
{\bf y'}=\begin{pmatrix}
1 & 0 & 1 \\
1 & 1 & 0 \\
1 & 1 & 1 \\ 
\end{pmatrix}
$$
where ${\bf y'}={\bf y} ~\hat{~}~ {\bf b}$.

\smallskip
\noindent
{\bf Remark:} Unlike removal-based unlearning, correction-based unlearning does not support irrecoverability, as each correction directly modifies class labels without accounting for prior corrections.

\subsection{Summary} \label{sec:advantage}

In this section, we summarize how our approach overcomes the limitations of the membership-proof-based method:

\smallskip
\noindent
{\bf Multi-Granular Unlearning.} Our bit-masking technique enables the selective removal of specific data units' impact on trained models so as to support multi-granular unlearning. The feature-level, sample-level, and class-level unlearning involve bit matrices with dimensions of $N \times J$, $N \times 1$ and $N \times K$ respectively. Moreover, a data owner can combine three bit matrices to simultaneously achieve a blend of three unlearning types within a single unlearning request.
    
\smallskip
\noindent
{\bf Invariable Datasets.} Our approach does not require continuous updates to the training datasets. The proofs can be consistently generated based on the initially committed dataset, eliminating the need to recompute commitments for updated datasets with each unlearning request. Additionally, the static nature of the dataset inherently prevents model trainers from covertly adding unauthorized data to the training datasets.

\smallskip
\noindent
{\bf Invariable Arithmetic Circuits.} Given a specific training dataset, the arithmetic circuits for constructing zero-knowledge proofs in our framework remain immutable to the addition or removal of data units' impact from model updates as these operations are governed by the bit matrix. This design ensures that any modifications to the dataset, such as adding new data or excluding unlearned data, do not alter underlying circuit structures. 

\smallskip
\noindent
{\bf Enhanced Privacy.} Our bit-masking approach enhances privacy by concealing both the indices and the quantity of unlearned data. Since the bit matrix is committed, it remains indistinguishable when bits are set to zero, and the Hamming weight of the matrix is kept confidential.

\smallskip
\noindent
{\bf Reduced Overheads.} Our bit-masking technique eliminates the need for computationally and communication-intensive hash-chain-based membership proofs.

\iffalse
CNNs allow multiple devices to collaboratively train a CNN model on decentralized data without sharing raw data, preserving privacy by only transmitting model updates rather than individual data points as shown in Figure~\ref{fig:fed-cnn}.
\fi

\section{Resistance to Forging Attacks} \label{sec:forgingattack}

We formally define a forgery game:

\smallskip
\noindent
{\bf Definition 1 (Forgery Game).} {\em In SGD-optimization, given a training dataset {\bf D} and a list of unlearned data {\bf U}, an adversary ${\mathcal{A}}$ wins the game if she or he can identify a minibatch ${\bf \tilde{d}} \subset {\bf D} \backslash {\bf U} \wedge |{\bf \tilde{d}}| \geq 1$ that generates similar gradient updates as a target minibatch ${\bf d} \subset {\bf D} \wedge {\bf d} \cap {\bf U} \neq \emptyset \wedge |{\bf d}| \geq 1$.}
\begin{equation}
   ||\nabla {\mathcal{L}}({\bf w}, {\bf d}) - \nabla {\mathcal{L}}({\bf w}, {\bf \tilde{d}})|| < \epsilon, ~\forall~\epsilon>0
\label{eqn:forgerygame} 
\end{equation}
where $||\cdot||$ is the Euclidean norm, $\wedge$ is the logical AND operator and $\epsilon$ is the {\em forgery error}. When $\epsilon = 0$, it is referred to as {\em exact forgery}, and when $\epsilon$ is a small positive value, it is termed {\em approximate forgery}. In such case, training a model over ${\bf \tilde{d}}$ can be seen as training it over ${\bf d}$ so that the impact of the unlearned data in ${\bf d}$ persists in the model. Notably, in our definition, $|{\bf \tilde{d}}|$ does not necessarily need to match $|{\bf d}|$, as the goal of forging attacks is to retain the impact of the unlearned data. Therefore, forging the gradient updates of any target minibatch containing the unlearned data can effectively fulfill this goal.

Two recent studies \cite{Thudi2021OnTN, zhang24h} propose different approaches, namely, {\em Adversary-Controlled Random Sampling} and {\em Replacement with Closest Class-Wise Neighbor} to carry out approximate forging attacks. We propose effective solutions to resist them, respectively.

\subsection{Adversary-Controlled Random Sampling}

\subsubsection{Problem Definition}

Aimed at MSGD optimization, the work \cite{Thudi2021OnTN} leverages random sampling, repeatedly generating minibatches ${\bf \tilde{d}} \subset {\bf D} \backslash {\bf U}$ and selecting the one with the smallest gradient distance to the target minibatch ${\bf d} \subset {\bf D}$. The root cause of such forging attack lies in the ability of model trainers to manipulate randomness, enabling them to strategically assemble data samples into minibatches. Note that each combination of data samples within a dataset corresponds to a gradient update. Thus, we begin with an important notation:

\smallskip
\noindent
{\bf Definition 2 (Search Space).} {\em The search space ${\mathcal{S}}$ for a given dataset refers to the set of gradient updates generated by all possible combinations of data samples.}

\smallskip
\noindent
Therefore, for a minibatch size $|{\bf \tilde{d}}|$, the adversary-controlled randomness leads to the search space dimensions for the forging minibatch ${\mathcal S}_f$ and the target minibatch ${\mathcal S}_t$ being $\binom{|{\bf D}| - |{\bf U}|}{|{\bf \tilde{d}}|}$ and $2^{|{\bf D}| - |{\bf U}|}$, respectively. The adversary's goal is to identify a minibatch from one of the $\binom{|{\bf D}| - |{\bf U}|}{|{\bf \tilde{d}}|}$ possible combinations that can replicate the gradient updates of an arbitrary-sized target minibatch including the unlearned data. Thus, this random sampling forging problem can be reframed as finding collisions between two large search spaces. Table \ref{tab:forgingspace} provides an illustrative comparison of forging spaces for various values of $|{\bf D}|$. Both search spaces are astronomically large, potentially resulting in a non-negligible number of pairs $\{{\bf \tilde{d}} \in {\mathcal{S}}_f, {\bf d} \in {\mathcal{S}}_t\}^*$ that satisfy the constraint in Eqn. \raf{eqn:forgerygame}.

\begin{table}[!ht]
\centering
\caption{An exemplar comparison of search spaces with varying $|{\bf D}|$, where $|{\bf \tilde{d}}|=10$ and $|{\bf U}|=1$.}
\resizebox{0.41\textwidth}{!}{%
\begin{tabular}{|c|c|c|c|c|}
\hline
|{\bf D}| & 256 & 512 & 768 & 1024  \\ \hline
$|{\mathcal{S}}_f|$ & 2.7E+17 & 3.1E+20 & 1.8E+22 & 3.3E+23 \\ \hline
$|{\mathcal{S}}_t|$ & 5.8E+76 & 6.7E+153 & 7.8E+230 & 9E+307 \\ \hline
\end{tabular}
}
\label{tab:forgingspace}
\end{table}

\subsubsection{Solution}

Thus, we propose incorporating {\em publicly verifiable randomness} into MSGD-optimized training to ensure that the randomness remains immune to manipulation by malicious model trainers. To support this approach, we have:

\smallskip
\noindent
{\bf Lemma 1.} {\em When minibatches in MSGD-optimized training are generated with publicly verifiable randomness, the dimension of the search space ${\mathcal{S}}_f$ for forging minibatches is reduced to $\lceil \frac{|{\bf D}|}{|{\bf \tilde{d}}|} \rceil$ at each epoch.}

\smallskip
\noindent
{\em Proof}. When minibatch sampling is beyond the adversary's control, the search space ${\mathcal{S}}_f$ is limited to $\lceil \frac{|{\bf D}|}{|{\bf \tilde{d}}|} \rceil$ minibatches. 

\smallskip
\noindent
{\bf Theorem 1.} {\em When minibatches in MSGD-optimized training are generated with publicly verifiable randomness, the probability of successfully executing (approximate) forging attacks is tremendously diminished.}

\smallskip
\noindent
{\em Proof.} With publicly verifiable randomness, we can see that the problem is reduced to matching the gradient updates of $\lceil \frac{|{\bf D}|}{|{\bf \tilde{d}}|} \rceil$ forging minibatches to those in the search space ${\mathcal{S}}_t$ for target minibatches. Let $p \in (0, 1)$ denote the probability of a single randomly sampled forging minibatch colliding with the target search space ${\mathcal{S}}_t$. By applying Bernoulli trials, we can conclude that for each epoch, the probability of at least one collision among $\binom{|{\bf D}| - |{\bf U}|}{|{\bf \tilde{d}}|}$ combinations is significantly greater than that among $\lceil \frac{|{\bf D}|}{|{\bf \tilde{d}}|} \rceil$ combinations due to $\binom{|{\bf D}| - |{\bf U}|}{|{\bf \tilde{d}}|} \gg \lceil \frac{|{\bf D}|}{|{\bf \tilde{d}}|} \rceil$:
\begin{equation*}
 \big(1-(1-p)^{\binom{|{\bf D}| - |{\bf U}|}{|{\bf \tilde{d}}|}}\big) \gg \big(1-(1-p)^{\lceil \frac{|{\bf D}|}{|{\bf \tilde{d}}|} \rceil}\big)   
 \label{eqn:probability}
\end{equation*}
\noindent
{\bf Remark:} A recent work (CCS \textquotesingle 23) \cite{unforgeability} demonstrated the infeasibility of exact forgery under specific conditions, in the case where adversaries have full control over the randomness. The work introduces a fast-to-check condition for assessing the forgeability by verifying whether there exist non-trivial solutions to a specific system of multivariate equations relevant to the MSGD optimizations. Proving the absolute probabilistic negligibility of $p$ for approximate forgery is more challenging, as it depends not only on factors such as the complexity of the model architecture and the size and distribution of the training dataset but also on the magnitude of the forgery error. Thus, we leave this issue for future research.

\subsubsection{Implementation}

To achieve this, we propose leveraging a well-established cryptographic primitive, {\em Verifiable Random Functions (VRFs)} \cite{VRF}, to securely and provably generate randomness for constructing minibatches, ensuring it remains beyond the influence of malicious model trainers. We begin with a brief introduction to VRFs. The unpredictability of randomness generated by VRFs ensures that malicious model trainers cannot strategically construct favorable minibatches to execute forging attacks.A model trainer can use a VRF-generated value $r$ to generate the index $i = r~\text{mod}~(|{\bf D}| - 1)$, which randomly selects a data sample from the training dataset. To ensure random sampling without replacement, the model trainer can repeat the sampling process until no duplicates are selected.

\smallskip
\noindent
{\bf Remark:} Note that a significant number of random values, independent of the unlearning process, can be generated during this randomness generation phase. This phase can be treated as a "pre-processing step" and performed at any point prior to the unlearning process. 

\subsection{Replacement With Class-Wise Neighbor}

\subsubsection{Problem Definition}

Alternatively, the study \cite{zhang24h} proposes constructing a minibatch for forgery by replacing the unlearned data $({\bf x}_u, y_u)$ with its nearest class-wise neighbor ${\mathcal{N}}({\bf x}_u, y_u) \in {\bf D} \backslash {\bf U}$ within any minibatch ${\bf d}$, where $({\bf x}_u, y_u) \in \bf d$:
\begin{equation}
    {\mathcal{N}}({\bf x}_u, y_u)=\text{argmin}_{({\bf x}, y) \in {\bf D} \backslash {\bf U}, y=y_u} ||{\bf x} - {\bf x}_u||
    \label{eqn:neighbour}
\end{equation}
While more efficient than random sampling, this method has limited applicability, as it is effective only when the closest class-wise neighbor happens to be a gradient replica of the unlearned data. Even so, using VRFs to generate publicly verifiable randomness does not mitigate this issue. Specifically, any randomly generated minibatch that includes such a neighbor, ${\mathcal{N}}({\bf x}_u, y_u)$, can replicate the model updates of a minibatch containing the unlearned data $({\bf x}_u, y_u)$. The presence of gradient replicas in the training dataset not only facilitates adversaries in forging gradient updates during MSGD optimization but also compromises the integrity of both SGD and BGD optimization processes. This is particularly concerning in the widely-used "sampling without replacement" approach, where all data samples are utilized during retraining. In such a setup, gradient replicas, if present in the dataset, are inevitably used for training. Even if the unlearned data is unlearned from the models, its gradient replicas may still persist, exerting influence on the models. Moreover, as indicated in Eqn. \raf{eqn:neighbour}, the "Class-Wise Neighbor Replacement" approach fails to consider cases where a data sample, despite not being closely related to the unlearned data in terms of distance, may still produce highly similar gradients.

\subsubsection{Detection}

Thus, we propose a robust detection mechanism to identify forging attacks stemming from gradient replicas across all three optimization methods. Specifically, model trainers must generate a zero-knowledge proof before each model update to demonstrate whether the distance between the gradient update produced by each data sample in the minibatch and that of the unlearned data falls below a predefined threshold $\xi$:
$$
||\nabla {\mathcal{L}} ({\bf x}_m, y_m) - \nabla {\mathcal{L}} ({\bf x}_u, y_u)|| \leq \xi,~ \forall m \in \{1, ..., |{\bf \tilde{d}}|\}
$$
Any data sample with a distance no greater than $\xi$ is considered a gradient replica to the unlearned data. This proof can be efficiently constructed using zero-knowledge proofs designed for arithmetic circuits.

\subsubsection{Solution} 

In this section, we propose a potential solution to resist such forging attacks following their detection. A direct approach to resist these attacks is to unlearn the identified gradient replicas, thereby preventing their continued influence on trained models. However, implementing this solution poses several challenges, including but not limited to, the following, requiring careful consideration:
\begin{itemize}
    \item{\bf Legal and ethical dilemmas}: This could raise disputes regarding intellectual property or data ownership, potentially infringing on the rights of the data owners associated with these gradient replicas.

    \item{\bf Risk of unlearning abuse}: Malicious data owners could exploit the unlearning mechanism to prevent legitimate data from contributing to model training, undermining the overall integrity and reliability of the model.
    
    \item{\bf Resource consumption and efficiency concerns}: Processing unlearning requests for gradient replicas may impose additional computational and storage overheads, particularly if the number of such requests is high. This may degrade system performance and scalability.
    
    \iffalse
    \item{\bf Impact on model generalization}: This could hinder the model’s ability to generalize well on unseen data.
    \fi
\end{itemize}
These challenges underscore the importance of carefully designing and implementing unlearning mechanisms to effectively address potential issues once forging attacks are identified. We leave this as an open direction for future research.
 
\subsection{Summary}

Effectively resisting forging attacks requires the integration of multiple techniques. On the one hand, our detection mechanism excels at identifying and mitigating forging attacks when gradient replicas are present, as these replicas can mimic the behavior of the unlearned data. On the other hand, in the absence of gradient replicas, adversaries must resort to an exhaustive search for a combination of data samples excluding the unlearned data that can approximate either the gradient updates of the unlearned data or those of a combination containing it. Detecting such forging attacks requires a computationally demanding search for collisions between the two search spaces, ${\mathcal{S}}_f$ and ${\mathcal{S}}_t$. 
Incorporating publicly verifiable randomness tremendously reduces the dimension of ${\mathcal{S}}_f$, significantly diminishing the feasibility of such forging attacks.

\section{Extensions} \label{sec:extension}

Our approach supports the following existing framework:

\smallskip
\noindent
{\bf SISA Unlearning} \cite{SISA} (Sharded, Isolated, Sliced, and Aggregated) is a powerful framework that improves unlearning efficiency by confining the influence of individual data units to localized portions of the training process. The dataset is partitioned into shards, each training a separate model instance to confine the impact of individual data units. Shards are further divided into slices, processed sequentially during training, allowing fine-grained control over data influence. The final model is an aggregation of shard-trained models, typically using ensemble learning. In this case, model trainers can simply furnish a zkPoT for the shard-trained models including the unlearned data.

\smallskip
\noindent
{\bf Federated environments} allow data providers to compute gradients locally on their own datasets. The model trainer then gathers and aggregates these local gradients to produce a global gradient, which is shared back with the data providers for model updates. This approach offers greater privacy for data providers compared to centralized setups. Unlike centralized environments, where model trainers must supply zkPoTs themselves, in federated settings, this responsibility shifts to the data providers.

\section{Experiments} \label{sec:experiment}

We conducted experiments using a representative instantiation of our framework with {\em Groth16} \cite{groth16}, a widely-used zkSNARK known for its efficient constant verification time and proof size, albeit requiring a trusted setup. Other zero-knowledge proof systems can also be utilized, depending on the performance requirements of specific scenarios. We utilized the {\em zkutil} \cite{zkutil} library, a Rust-based toolkit to handle arithmetic circuits on the {\em BN256} elliptic curve of a 256-bit prime-order group. We used the popular {\em Circom} \cite{circom} compiler, along with its machine learning extension library {\em CircomLib-ML} \cite{circomml}, to construct zkSNARK circuits. All experiments were performed on a Ubuntu system with a 13th Gen Intel Core i5 processor (2500 MHz, 16 GB RAM).

We evaluated the performance of zkPoTs for multi-granular unlearning and forging attack detection using MSGD optimization during a single training step across varying batch sizes of 20, 30, 40, and 50 samples to demonstrate scalability. A single training step offers a more indicative and representative measure than full training, as the latter's performance can vary significantly depending on the size of the training dataset. With Groth16, proving complexity increases linearly with the number of steps, while verification complexity and proof size always remain constant. Additionally, we employed two classic machine learning models: {\em linear regression (LR)} and {\em neural network (NN)} for a multi-label classification task.

\subsection{R1CS Constraints}

We performed an ablation study to assess the performance impact of integrating our bit-masking technique by comparing the original zkPoT instantiations with their bit-masked counterparts\footnote{Directly comparing with the membership-based approach \cite{eisenhofer2023} is not appropriate, as their performance was evaluated using BGD optimization under completely different settings, with significantly higher costs than ours. Our ablation study, however, provides valuable insights into the efficiency and scalability of our bit-masking technique.}. We used R1CS (Rank-1 Constraint System) metric, commonly employed in zkSNARKs to represent computations as a set of mathematical constraints. The number of R1CS constraints reflects the computational complexity of a certain arithmetic circuit, independent of underlying zero-knowledge proof systems.

\begin{figure}[!ht]
	\centering
	\includegraphics[width=0.36\textwidth]{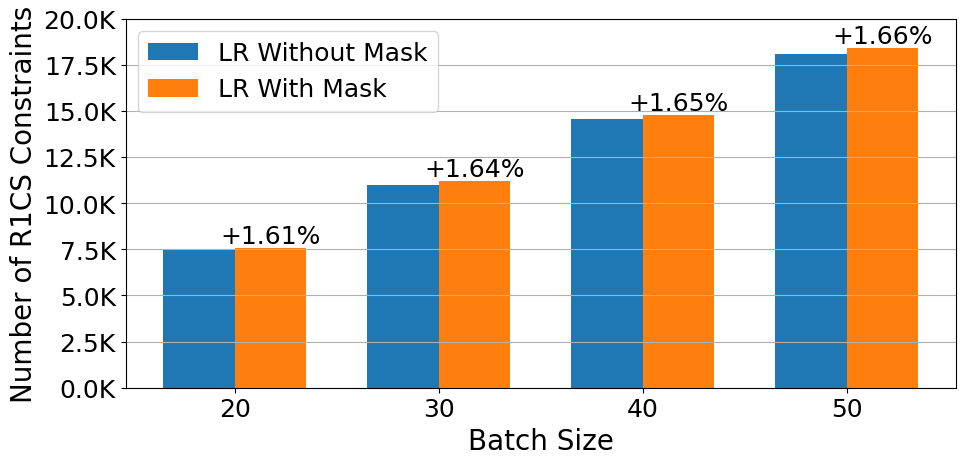}
	\caption{R1CS constraints in a linear regression model.}
	\label{fig:linear_regression}
\end{figure}

\smallskip
\noindent
{\bf Feature-Level Unlearning.} Figure \ref{fig:linear_regression} compares the number of R1CS constraints for a linear regression model with 4 features. It can be observed that the R1CS complexity grows linearly with increased batch sizes. Notably, incorporating our bit-masking technique resulted in only a small increase of approximately 1.6\% in the number of R1CS constraints, as the overall computation is dominated by commitment generation, with bit-masking contributing only a minor increase.

\begin{figure}[!ht]
	\centering
	\includegraphics[width=0.36\textwidth]{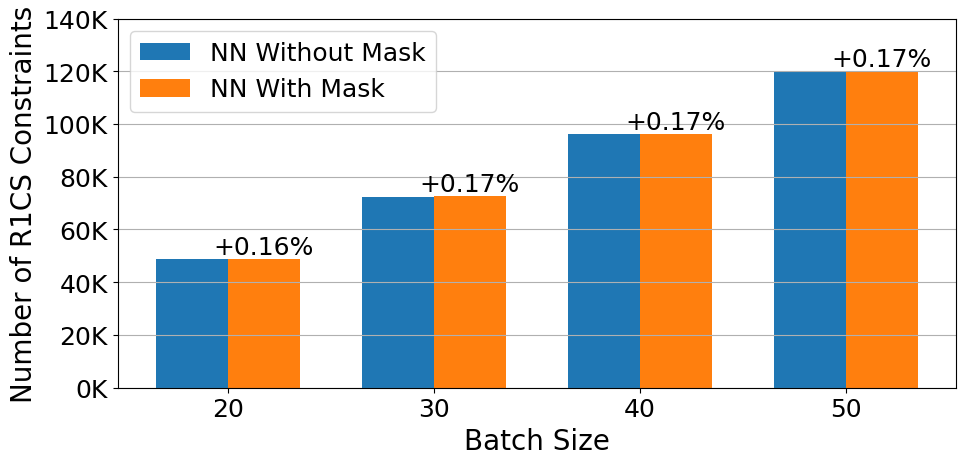}
	\caption{R1CS constraints in a neural network model.}
	\label{fig:neural_network}
\end{figure}

\smallskip
\noindent
{\bf Feature-Level \& Class-Level Unlearning.} We applied both feature-level and class-level unlearning to a 3-layer neural network with 4 input features, 4 hidden neurons, and 4 output labels. Figure \ref{fig:neural_network} shows a comparison of R1CS constraint counts. Likewise, the results demonstrate that incorporating our bit-masking technique led to only a minimal 0.17\% increase in the number of R1CS constraints.

\begin{figure}[!ht]
	\centering
	\includegraphics[width=0.36\textwidth]{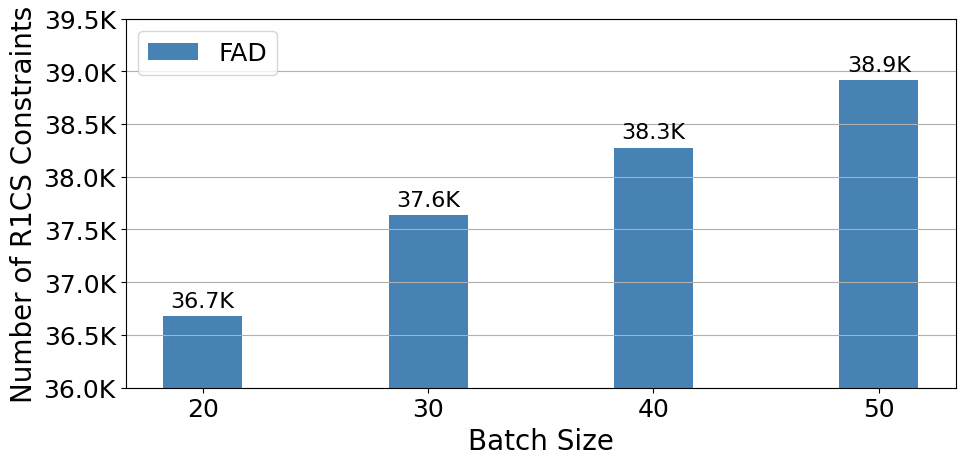}
	\caption{R1CS constraints for forging attack detection in a linear regression model.}
	\label{fig:forging_attack}
\end{figure}

\smallskip
\noindent
{\bf Forging Attack Detection (FAD).} We measured the number of R1CS constraints required for detecting forging attacks, as shown in Figure~\ref{fig:forging_attack}. While the number of R1CS constraints increases linearly with batch size, the increase is much smaller than those in Figures \ref{fig:linear_regression} and \ref{fig:neural_network}, reaching only the hundreds, indicating that our detection mechanism is highly efficient, making it practical for real-world applications.

\subsection{Running Time \& Proof Size}

We present another two key metrics of our instantiation, running time and proof size for bit-masked zkPoTs. Unlike R1CS constraints, these two metrics are provided for reference, as they are heavily influenced by the underlying zero-knowledge proof implementations and the hardware on which they are executed. 

\smallskip
\noindent
{\bf Running Time.} Table \ref{tab:runningtime} presents the running times in seconds, for the largest 50-sample minibatch. As shown, the proving times are all under a single second, making them practically applicable. Notably, the verification time remains constant at 0.2 seconds, while the trusted-setup generation time is 8.9 seconds, more than 10$\times$ the proving time. This highlights one of the limitations of the membership-proof-based approach, which requires a time-consuming trusted setup for each unlearning request. 

\smallskip
\noindent
{\bf Proof Size.} Moreover, the proof size of our Groth16 instantiation is small and remains constant at 192 bytes.

\begin{table}[!ht]
\centering
\caption{The running time in seconds with a batch size of 50.}
\resizebox{0.43\textwidth}{!}{%
\begin{tabular}{|c|c|c|c|c|}
\hline
\textbf{\begin{tabular}[c]{@{}c@{}}LR\\ Prover\end{tabular}} & \textbf{\begin{tabular}[c]{@{}c@{}}NN\\ Prover\end{tabular}} & \textbf{\begin{tabular}[c]{@{}c@{}}FAD\\ Prover\end{tabular}} & \textbf{\begin{tabular}[c]{@{}c@{}}Verifier\\ For All\end{tabular}} & \textbf{\begin{tabular}[c]{@{}c@{}}Trusted Setup\\ Generation\end{tabular}} \\ \hline
0.37 & 0.94  & 0.62 & 0.2  & 8.9                                                                         \\ \hline
\end{tabular}
}
\label{tab:runningtime}
\end{table}

\section{Conclusion}

In this work, we introduced zkUnlearner, the first zero-knowledge framework for multi-granular and forgery-resistant verifiable unlearning. Our approach leverages a bit-masking technique to enable efficient and privacy-preserving unlearning at various granularities, addressing key limitations of existing methods. Additionally, we proposed effective solutions to resist forging attacks, enhancing the framework's robustness. In the future, we aim to focus on exploring verifiable approximate unlearning and developing stronger defenses against forging attacks.

\bibliographystyle{ACM-Reference-Format}

\bibliography{reference}

\end{document}